%% file: SUSY-2016-02.tex
\pdfoutput=1
\newcommand*{\ATLASLATEXPATH}{}
\documentclass[cernpreprint,texlive=2011,txfonts,UKenglish,paper=a4]{atlasdoc}

\usepackage{atlaspackage}
\usepackage{atlasbiblatex}

\usepackage{atlascontribute}

\usepackage{atlasphysics}

\usepackage{xspace}
\usepackage{rotating}
\usepackage{multirow}

\newcommand{\pythia} {{\sc Pythia}\xspace}

\newcommand{\madgraph} {{MG5\_aMC@NLO}\xspace}

\newcommand{\geant} {{\sc Geant4}\xspace}

\newcommand{\rhad} {\mbox{$R$-hadron}\xspace}
\newcommand{\rhads} {\mbox{$R$-hadrons}\xspace}

\newcommand{\lumi} {3.2~\ifb\xspace}
\newcommand{\lumiUnc}{5\%\xspace}

\renewcommand{\pt} {\ensuremath{p_{\mathrm{T}}}\xspace}

\newcommand{\zmumu} {\ensuremath{Z \to \mu\mu}\xspace}
\newcommand{\etmiss} {\ensuremath{E_{\mathrm{T}}^\mathrm{miss}}\xspace}
\newcommand{\ptmiss} {\ensuremath{\vec{p}_{\mathrm{T}}^\mathrm{~miss}}\xspace}

\newcommand{\pb} {\ensuremath{\mathrm{{pb}}}\xspace}
\renewcommand{\MeV} {\ensuremath{\mathrm{{Me\kern -0.1em V}}}\xspace}
\renewcommand{\GeV} {\ensuremath{\mathrm{{Ge\kern -0.1em V}}}\xspace}
\renewcommand{\TeV} {\ensuremath{\mathrm{{Te\kern -0.1em V}}}\xspace}

\newcommand{\MeVgcm} {\ensuremath{\MeV\mathrm{g}^{-1}\mathrm{cm}^{2}}\xspace}
\newcommand{\dedx} {{\ensuremath{\mathrm{d}E/\mathrm{d}x}}\xspace}

\newcommand{\limitExpGluinoStable} {{1655~\GeV}\xspace}
\newcommand{\limitGluinoStable} {{1580~\GeV}\xspace}
\newcommand{\limitExpSbottomStable} {{865~\GeV}\xspace}
\newcommand{\limitSbottomStable} {{805~\GeV}\xspace}
\newcommand{\limitExpStopStable} {{945~\GeV}\xspace}
\newcommand{\limitStopStable} {{890~\GeV}\xspace}

\AtlasTitle{Search for heavy long-lived charged \rhads with the ATLAS detector in \lumi of proton--proton collision data at $\sqrt{s} = 13~\TeV$}
\author{The ATLAS Collaboration}
\AtlasRefCode{SUSY-2016-02}
\PreprintIdNumber{CERN-EP-2016-131}
\AtlasJournalRef{Physics Letters B (2016), pp. 647-665}
\AtlasDOI{10.1016/j.physletb.2016.07.042}
\AtlasAbstract{%
  A search for heavy long-lived charged \rhads is reported using a data sample corresponding to \lumi of proton--proton collisions at $\sqrt{s} = 13~\TeV$ collected by the ATLAS experiment at the Large Hadron Collider at CERN. The search is based on observables related to large ionisation losses and slow propagation velocities, which are signatures of heavy charged particles travelling significantly slower than the speed of light. No significant deviations from the expected background are observed. Upper limits at 95\% confidence level are provided on the production cross section of long-lived \rhads in the mass range from 600~\GeV to 2000~\GeV and gluino, bottom and top squark masses are excluded up to \limitGluinoStable, \limitSbottomStable and \limitStopStable, respectively.
}

\hypersetup{pdftitle={ATLAS draft},pdfauthor={The ATLAS Collaboration}}

\begin{document}

\maketitle



\section{Introduction}
\label{sec:intro}

Heavy long-lived particles (LLP) are predicted in a range of theories extending the Standard Model (SM) in an attempt to address the hierarchy problem \cite{Susskind:1982mw}. These theories include supersymmetry (SUSY)~\cite{Golfand:1971iw,Volkov:1973ix,Wess:1974tw,Wess:1974jb,Ferrara:1974pu,Salam:1974ig}, which allows for long-lived charged sleptons ($\tilde{\ell}$), squarks ($\tilde{q}$), gluinos ($\tilde{g}$) and charginos ($\tilde{\chi}^{\pm}_{1}$) in models that either violate~\cite{Dreiner:1997uz,Berger:2003kc,Barbier:2004ez} or conserve~\cite{Farrar:1978xj,Fairbairn:2006gg,Kolda:1997wt,Baer:1998pg,Gates:1999ei,Giudice:2004tc,ArkaniHamed:2004yi} $R$-parity.

Heavy long-lived charged particles can be produced at the Large Hadron Collider (LHC). A search for composite colourless states of squarks or gluinos together with SM quarks or gluons, called \rhads~\cite{Farrar:1978xj}, is presented in this Letter. The search exploits the fact that these particles are expected to propagate with a velocity, $\beta=v/c$, substantially lower than one and to exhibit a specific ionisation energy loss, \dedx, larger than that for any charged SM particle. Similar searches have been performed previously by the ATLAS and CMS Collaborations~\cite{ATLAS:2014fka,Chatrchyan:2013oca} using data samples from Run~1 at the LHC. No excesses of events above the expected backgrounds were observed, and lower mass limits were set at 95\% confidence level (CL) around 1300~\GeV for gluino \rhads.

\rhads can be produced in $pp$ collision as either charged or neutral states, and can be modified to a state with different charge by interactions with the detector material~\cite{Kraan:2004tz,Mackeprang:2009ad}, arriving as neutral, charged or doubly charged particles in the muon spectrometer (MS) of the ATLAS detector. This search does not use information from the MS and follows the ``MS-agnostic'' \rhad search approach in Ref.~\cite{ATLAS:2014fka}. This strategy avoids assumptions about \rhad interactions with the detector, especially in the calorimeters, and is sensitive to scenarios in which \rhads decay or become neutral (via parton exchange with the detector material in hadronic interactions) before reaching the MS.

\section{ATLAS detector}
\label{sec:detector}

The ATLAS detector~\cite{PERF-2007-01} is a multi-purpose particle-physics detector consisting of an inner detector (ID) immersed in an axial magnetic field to reconstruct trajectories of charged particles, calorimeters to measure the energy of particles that interact electromagnetically or hadronically and a MS within a toroidal magnetic field to provide tracking for muons. With near $4\pi$ coverage in solid angle,\footnote{ATLAS uses a right-handed coordinate system with its origin at the nominal interaction point (IP) in the centre of the detector and the $z$-axis coinciding with the axis of the beam pipe. The $x$-axis points from the IP to the centre of the LHC ring, and the $y$-axis points upward. Cylindrical coordinates ($r$, $\phi$) are used in the transverse plane, $\phi$ being the azimuthal angle around the beam pipe. The pseudorapidity is defined in terms of the polar angle $\theta$ as $\eta = - \ln \tan(\theta/2)$.} the ATLAS detector is able to deduce the missing transverse momentum, \ptmiss, associated with each event. The components of particular importance to this search are described in more detail below.


The ID consists of two distinct silicon detectors and a straw tracker, which jointly provide good momentum measurements for charged tracks. The innermost part of the ID, a silicon pixel detector, typically provides four or more precision measurements for each track in the region $|\eta|<2.5$ at radial distances $3.4<r<13$~cm from the LHC beam line. All pixel layers are similar, except the innermost Insertable B-Layer (IBL)~\cite{Capeans:1291633}, which has a smaller pixel size and a reduced thickness, but also 4-bit instead of 8-bit encoding and hence poorer charge resolution than the other pixel layers. The charge released by the passage of a charged particle is rarely contained within a single silicon pixel, and a neural network algorithm~\cite{Aad:2014yva} is used to form clusters from the single pixel charges. For each cluster in the pixel detector a \dedx estimate can be provided, from which an overall \dedx measurement is calculated as a truncated mean to reduce the effect of the tail of the Landau distribution, by disregarding the one or two largest measurements~\cite{pixel_paper}. Radiation sensitivity of the IBL electronics results in the measured \dedx drifting over time. This effect is corrected by applying a dedicated time-dependent ionisation correction of 1.2\% on average. The mean and RMS of the \dedx measurement for a minimum-ionising particle are 1.12~\MeVgcm and 0.13~\MeVgcm, respectively, while the distribution extends to higher \dedx values, due to the remnants of the Landau tail.

\begin{figure}[tbp]
\centering
\includegraphics[width=0.98\textwidth]{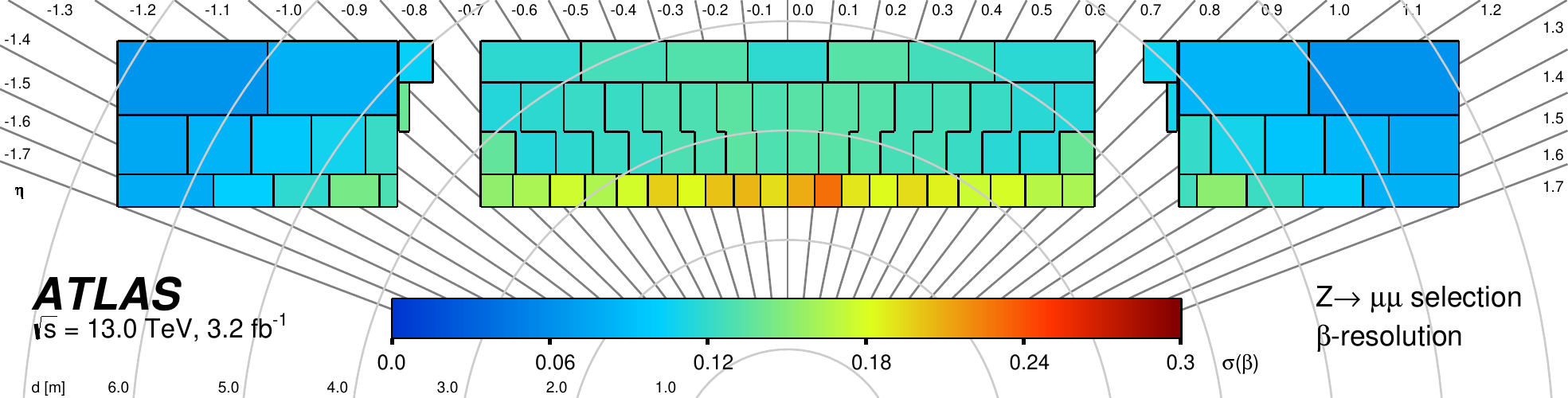}
\caption{Resolution of $\beta$ for different cells in the ATLAS tile calorimeter obtained from a \zmumu selection in data. The final $\beta$ measurement is a weighted average of the $\beta$ measurements in the cells traversed by the candidate.}
\label{fig:calo_betares}
\end{figure}

The ATLAS calorimeter in the central detector region consists of an electromagnetic liquid-argon calorimeter followed by a hadronic tile calorimeter.
The estimation of $\beta$ from time-of-flight measurements relies on timing and distance information from tile-calorimeter cells crossed by the extrapolated candidate track in three radial layers in the central barrel as well as an extended barrel on each side, as illustrated in Figure~\ref{fig:calo_betares}. To reduce effects of detector noise, only cells in which the associated particle has deposited a minimum energy $E_\text{min,cell}=500~\MeV$ are taken into account. The time resolution depends on the energy deposited in the cell and also the layer type and thickness of the cell.

A series of calibration techniques is applied to achieve optimal performance, using a \zmumu control sample. Muons on average deposit slightly less energy than expected from signal, but variations sufficiently cover the relevant range. First, a common time shift is applied for each short period of data taking (run) followed by five additional cell-by-cell $\beta$ corrections. A geometry-based cell correction is introduced to minimise the $\eta$ dependence of $\beta$ within each individual cell. This is done by taking into account the actual trajectory ($\eta$ and path length) of the extrapolated track in each calorimeter cell, to recalculate the distance-of-flight, instead of using the centre of the cell, as done in previous ATLAS searches (e.g.\ in~\cite{ATLAS:2014fka}). The effect is most prominent at the edges of the largest cells at high $|\eta|$ with shifts of up to 0.05 in $\beta$, and almost negligible for the cells at low $|\eta|$. An additional correction, linear in $|\eta|$ and only applied in simulation, is added to account for a timing mismodelling due to an imperfect simulation. This correction is again most prominent for the cells at high $|\eta|$ with shifts up to 0.1 in $\beta$. The Optimal Filtering Algorithm (OFA)~\cite{Fullana:2005dwa} used for the readout of the tile calorimeter cells is optimised for in-time signals and introduces a bias towards lower values of $\beta$ in the measured cell time of late-arriving particles. To compensate for this bias for late-arriving particles, a correction is estimated from a fit to simulated late signals. Cell times larger than 25~ns are discarded, to limit the size of the required correction. The size of the correction is up to 0.05 in $\beta$. A cell-time smearing is applied to adjust the cell-time resolution in simulation to that observed in data. The uncertainty in the single $\beta$ measurements is scaled up by about 12\%, based on the requirement that the pull distribution $(\beta - \beta_{\mathrm{true}}) / \sigma_{\beta}$ be a unit Gaussian. Finally the $\beta$ associated with the particle is estimated as a weighted average, using the $\beta$ measurement in each traversed cell and its uncertainty, $\sigma_{\beta}$.

After all calibrations, the single cell-time resolution ranges from 1.3~ns in cells at large radii to 2.5~ns in cells at small radii. The distances from the nominal interaction point (IP) to the cell centres are 2.4~m to 3.6~m (4.2~m to 5.7~m) at $|\eta| \sim 0$ ($|\eta| \sim 1.25$). This in turn results in a resolution of 0.06 to 0.23 in $\beta$, as shown in Figure~\ref{fig:calo_betares}. The larger cells at large radii have a better resolution due to the higher energy deposits and their increased distance from the IP.

\begin{figure}[tbp]
\centering
\includegraphics[width=0.48\textwidth]{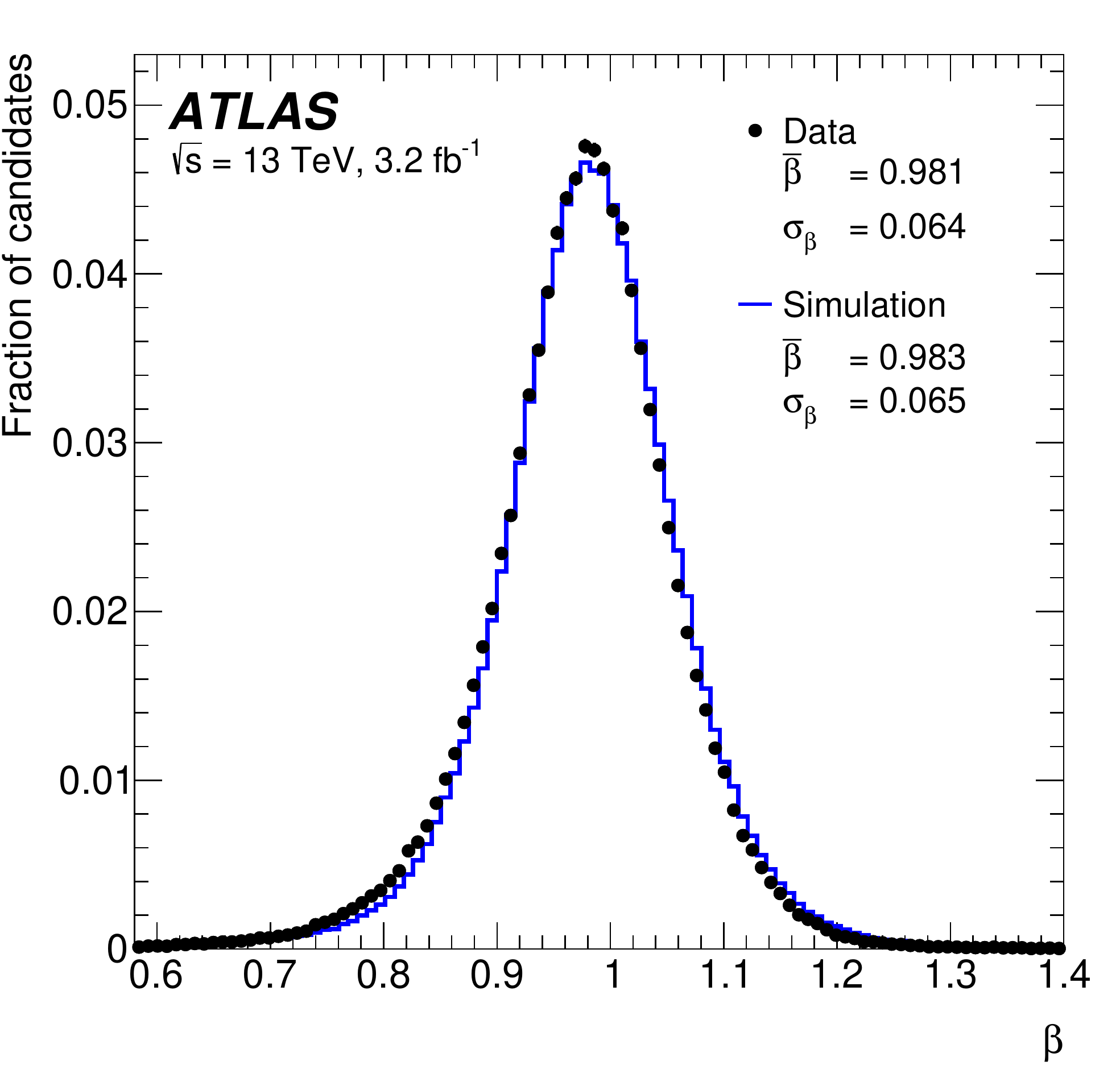}
\caption{Distributions of $\beta$ for data and simulation after a \zmumu selection. The values given for the mean and width are taken from Gaussian functions matched to data and simulation.}
\label{fig:calo_beta}
\end{figure}

As described in Section~\ref{sec:background}, the expected $\beta$ distribution for the background is determined from data. However, the $\beta$ distribution for the \rhad signal is obtained from simulation. Figure~\ref{fig:calo_beta} shows the $\beta$ distributions obtained for both data and simulation for a control sample of \zmumu events that is used to validate the $\beta$ measurement. Good agreement between data and simulation supports the use of the simulation to predict the behaviour expected for the \rhad signal.

\section{Data and simulated events}
\label{sec:samples}

The work presented in this Letter is based on \lumi of $pp$ collision data collected in 2015 at a centre-of-mass energy $\sqrt{s}=13~\TeV$. Reconstructed \zmumu events in data and simulation are used for timing resolution studies. Simulated signal events are used to study the expected signal behaviour.

\rhad signal events are generated with gluino (bottom-squark and top-squark) masses from 600~\GeV to 2000~\GeV (600~\GeV to 1400~\GeV). Pair production of gluinos and squarks is simulated in \pythia 6.427~\cite{Sjostrand:2006za} with the AUET2B~\cite{ATL-PHYS-PUB-2011-014} set of tuned parameters for the underlying event and the CTEQ6L1~\cite{ATLAS_PDF_set} parton distribution function (PDF) set, incorporating \pythia-specific specialised hadronisation routines~\cite{hadroutines,Kraan:2004tz,Farrar:2010ps} to produce final states containing \rhads. The masses of the other SUSY particles are set to very high values to ensure that their contribution to the production cross section is negligible. For a given sparticle mass the production cross section for gluino \rhads is typically an order of magnitude higher than for bottom-squark and top-squark \rhads. The probability for a gluino to form a gluon--gluino bound state is assumed, based on a colour-octet model, to be 10\%~\cite{Fairbairn:2006gg}. The associated hadronic activity produced by the colour field of the sparticle typically only possesses a small fraction of the initial energy of the sparticle~\cite{Fairbairn:2006gg}, which should therefore be reasonably isolated.

To achieve a more accurate description of QCD radiative effects, the \pythia events are reweighted to match the transverse-momentum distribution of the gluino--gluino or squark--squark system to that obtained in dedicated \madgraph v2.2.3.p0~\cite{Alwall:2014hca} events, as \madgraph can produce additional QCD initial-state radiation (ISR) jets as part of the hard process, while \pythia only includes showering to add jets to the event.

All events pass through a full detector simulation~\cite{Aad:1267853}, where interactions with matter are handled by dedicated \geant~\cite{Agostinelli:2002hh} routines based on different scattering models: the model used to describe gluino (squark) \rhad interactions is referred to as the generic (Regge) model~\cite{Mackeprang:2009ad}. The \rhads interact only moderately with the detector material, as most of the \rhad momentum is carried by the heavy gluino or squark, which has little interaction cross section. Typically, the energy deposit in the calorimeters is less than 10~\GeV.

All simulated events include a modelling of contributions from pile-up by overlaying minimum-bias $pp$ interactions from the same (in-time pile-up) and nearby (out-of-time pile-up) bunch crossings, and are reconstructed using the same software used for collision data. Simulated events are reweighted so that the distribution of the expected number of collisions per bunch crossing matches that of the data.

\section{Event selection}
\label{sec:selection}

Events are selected online via a trigger based on the magnitude of the missing transverse momentum, \etmiss. Large \etmiss values are produced mainly when QCD initial-state radiation (ISR) boosts the \rhad system, resulting in an imbalance between ISR and \rhads whose momenta are not fully accounted for in the \etmiss calculation. In particular, the adopted trigger imposes a threshold of 70~\GeV on \etmiss calculated solely from energy deposits in the calorimeters~\cite{ATLAS:trigger}. The signal efficiency of the \etmiss trigger varies between 32\% and 50\%, depending on the mass and type of the \rhad.

The offline event selection requires all relevant detector components to be fully operational; a primary vertex (PV) built from at least two well-reconstructed charged-particle tracks, each with a transverse momentum, \pt, above 400~\MeV; and at least one \rhad candidate track that meets the criteria specified below.

\rhad candidates are based on ID tracks with $\pt>50~\GeV$ and $|\eta|<1.65$. Candidates must not be within $\Delta R = \sqrt{\left( \Delta \eta \right)^2 + \left(\Delta \phi \right)^2} = 0.3$ of any jet with $\pt > 50~\GeV$, reconstructed using the anti-$k_t$ jet algorithm~\cite{jetalgo} with radius parameter set to 0.4. Furthermore, the candidates must not have any additional nearby ($\Delta R < 0.2$) tracks with $\pt > 10~\GeV$. Tracks reconstructed with $p>6.5~\TeV$ are rejected as unphysical. To ensure a well reconstructed track, a minimum number of seven hits in the silicon detectors is required. Of these, at least two clusters used to measure \dedx in the pixel detector are required, to ensure a good \dedx measurement. Candidates with $|z_0^\text{PV}\sin(\theta)| > 0.5~\text{mm}$ or $|d_0|>2.0~\text{mm}$ are removed, where $d_{0}$ is the transverse impact parameter at the candidate track's point of closest approach to the IP and $z_0^\text{PV}$ is the $z$ coordinate of this point relative to the PV. To suppress background muons stemming from cosmic-ray interactions, candidates with direction $\left(\eta, \phi\right)$ are rejected if an oppositely-charged track with almost specular direction, i.e.\ with $|\Delta\eta| < 0.005$ and $|\Delta\phi| < 0.005$ with respect to $\left( -\eta, \pi -\phi\right)$, is identified on the opposite side of the detector. In order to minimise the background from \zmumu decays, candidates are rejected if they result in an invariant mass closer than 10~\GeV to the mass of the $Z$ boson when combined with the highest-\pt muon candidate in the event. In addition to the above mentioned track-quality criteria, candidates must also satisfy observable-quality criteria, defined by an unambiguous $\beta\gamma$~determination from the \dedx value, estimated using an empirical relation (more details can be found in Ref.~\cite{ATLAS-CONF-2011-016}), determined from low-momentum pions, kaons and protons~\cite{ATLAS-CONF-2011-016}, and a $\beta$~measurement, with an uncertainty $\sigma_{\beta}$ of less than 0.12. In the following, $\beta\gamma$ refers to quantities derived from the \dedx measurement in the silicon pixel detector and $\beta$ refers to the time-of-flight-based measurement in the tile calorimeter.

After this initial selection, 226107 of the approximately 36~million initially triggered data events as well as 10\% to 15\% of simulated signal events (the percentage increases with hypothesised mass) remain. Only the candidate with the largest \pt is used in events with multiple \rhad candidates. The final signal selection, requiring a momentum above 200~\GeV as well as criteria summarised in Table~\ref{tab:rhad_ma_finalselection}, is based on $\beta\gamma$ and $\beta$, requiring $\beta\gamma<1.35$ ($<1.15$) for \rhad masses up to (greater than) 1.4~\TeV and $\beta < 0.75$ in all cases. The signal region is defined in the $m_{\beta\gamma}$--$m_\beta$ plane for each \rhad mass point, where $m_{\beta\gamma}$ and $m_\beta$ are extracted independently from the measurement of the momentum as well as $\beta\gamma$ and $\beta$, respectively, via $m=p/{\beta\gamma}$. The minimum mass requirements, $m_{\beta\gamma}^\text{min}$ and $m_\beta^\text{min}$, are set to correspond to a value about $2\sigma$ below the nominal \rhad mass value, given the mass resolution expected for the signal.

\begin{table}[btp]
  \small
  \centering
  \begin{tabular*}{\textwidth}{r| @{\extracolsep{\fill}} rrrrrrrr}
    \hline
    \hline
    Simulated   & & & & & & & & \\
    \rhad       &   600 &   800 &  1000 &  1200 &  1400 &  1600 &  1800 &  2000 \\
    mass [\GeV] & & & & & & & & \\
    \hline
    $\beta\gamma^\text{max}$     &  1.35 &  1.35 &  1.35 &  1.35 &  1.35 &  1.15 &  1.15 &  1.15 \\
    $\beta^\text{max}$           &  0.75 &  0.75 &  0.75 &  0.75 &  0.75 &  0.75 &  0.75 &  0.75 \\
    $m_{\beta\gamma}^\text{min}$ &   350 &   450 &   500 &   575 &   650 &   675 &   750 &   775 \\
    $m_\beta^\text{min}$         &   350 &   450 &   500 &   575 &   650 &   675 &   750 &   775 \\
    \hline
    \hline
  \end{tabular*}
  \caption{Final selection requirements as a function of the simulated \rhad mass.}
  \label{tab:rhad_ma_finalselection}
\end{table}
The total selection efficiency depends on the sparticle mass and varies between 9\% and 15\% for gluino and top-squark \rhads and 6\% to 8\% for bottom-squark \rhads. The lower efficiency for bottom squarks is expected, as \rhads are most likely produced in mesonic states, where those with down-type squarks tend to be neutral more often than those with up-type squarks, due to light-quark production ratios of $u:d:s \approx 1:1:0.3$~\cite{Fairbairn:2006gg} during hadronisation. The expected signal yield and efficiency, estimated background and observed number of events in data for the full mass range after the final selection are summarised in Table~\ref{tab:rhad_ma_finalnumbers}.

\section{Background estimation}
\label{sec:background}

\begin{figure}[tbp]
  \centering
  \includegraphics[width=.48\textwidth]{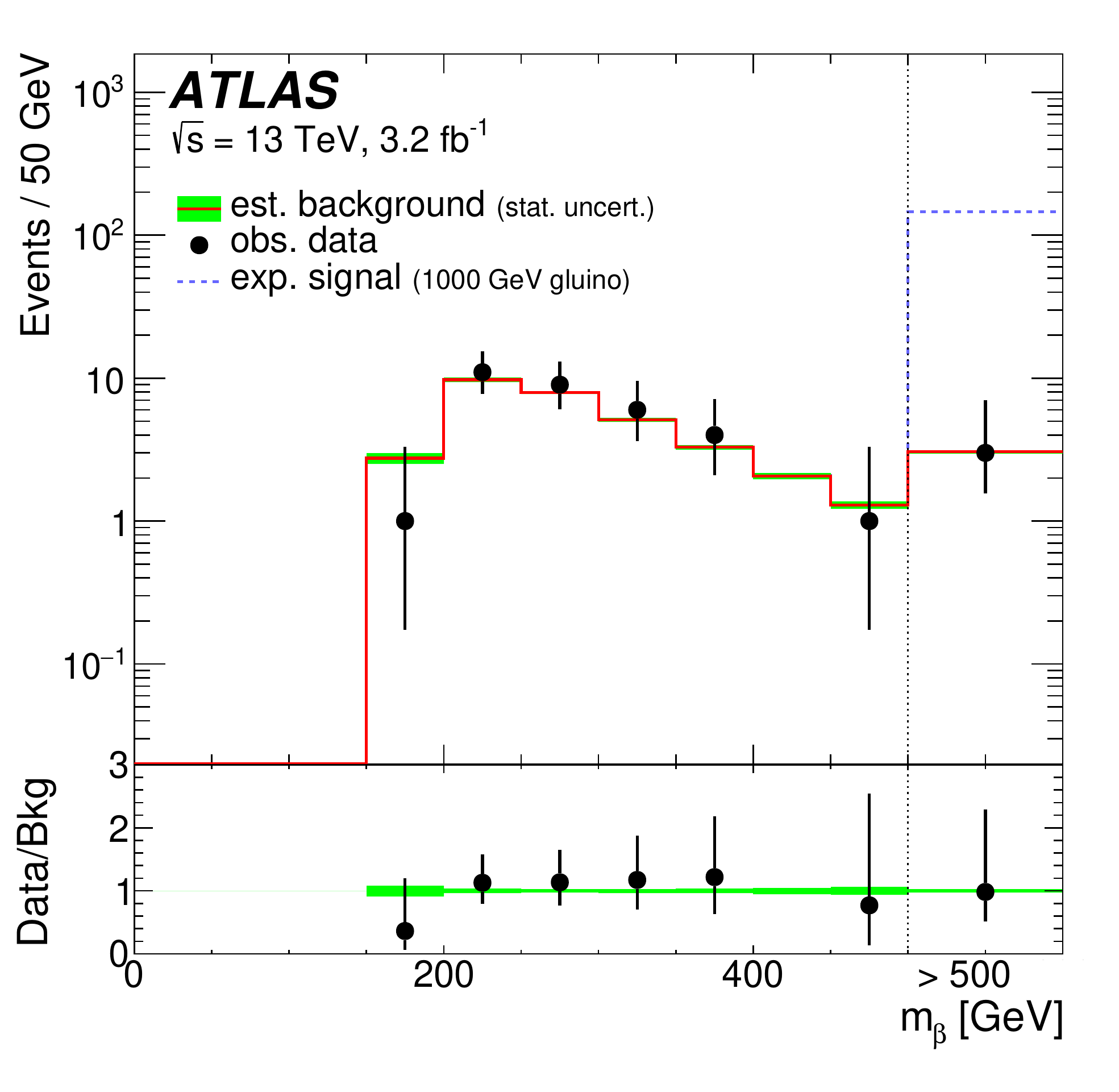}
  \includegraphics[width=.48\textwidth]{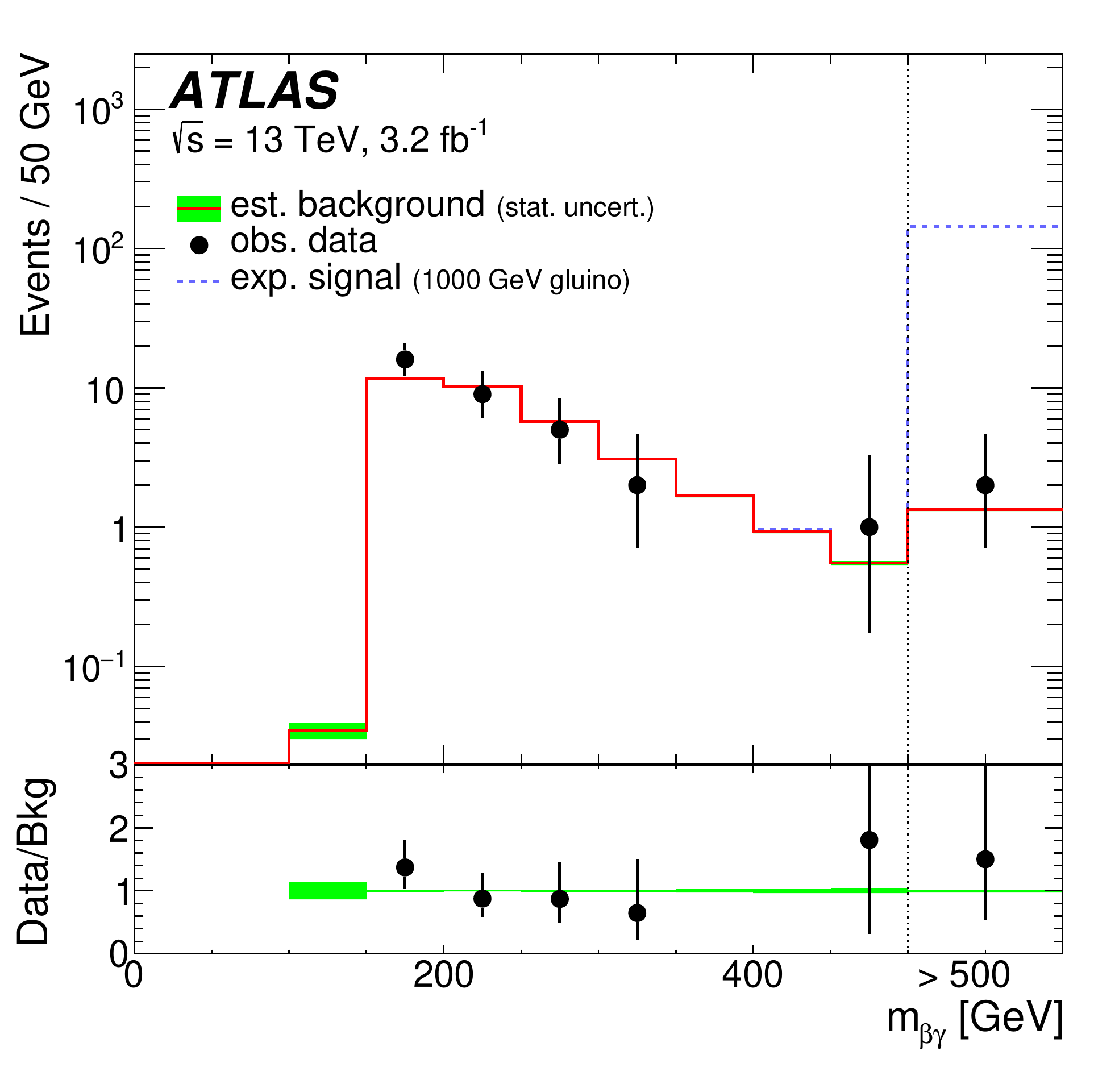}\\
  \caption{Data (black dots) and background estimates (red solid line) for $m_{\beta}$ (left) and $m_{\beta\gamma}$ (right) for the gluino \rhad search (1000~\GeV). The green shaded band illustrates the statistical uncertainty of the background estimate. The blue dashed lines illustrate the expected signal (on top of background) for the given \rhad mass hypothesis. The black dashed vertical lines at 500~\GeV show the mass selection and the last bin includes all entries/masses above. (For interpretation of the references to colour in this figure legend, the reader is referred to the web version of this article.)}
  \label{fig:rhadmsagnobkgestmass1000}
\end{figure}

The background is evaluated in a data-driven manner. First, probability distribution functions (pdf) in the momentum, and also in the $\beta$ and $\beta\gamma$ values, are determined from data. These pdfs are produced from candidates in data, which have passed the initial selection mentioned earlier, but fall in sidebands of the signal region, as described below. Background distributions in $m_\beta$ and $m_{\beta\gamma}$ are obtained by randomly sampling the pdfs derived above and then using the equation $m=p/{\beta\gamma}$. These mass distributions, which are normalised to the data events outside the signal region (i.e.\ not passing both mass requirements of the hypothesis in question), are shown in Figure~\ref{fig:rhadmsagnobkgestmass1000} along with the data and expected signal for the 1000~\GeV gluino \rhad mass hypothesis.

Each \rhad mass hypothesis has a different selection, and therefore corresponding individual background estimates are produced accordingly. The momentum pdf is produced from events that pass the momentum cut, but fail the $\beta$ and $\beta\gamma$ requirements in Table~\ref{tab:rhad_ma_finalselection} for the chosen \rhad mass hypothesis, but nonetheless have $\beta < 1$ and $\beta\gamma < 2.5$. The $\beta$ and $\beta\gamma$ pdfs are produced by selecting events which pass the respective $\beta$ and $\beta\gamma$ selection and have momentum in the range $50~\GeV < p < 200~\GeV$. Since momentum is correlated with $|\eta|$, any correlation between $|\eta|$ and $\beta$ ($\beta\gamma$) will lead to a correlation between momentum and $\beta$ ($\beta\gamma$), invalidating the background estimate. The size and impact of such correlations are reduced by determining the three pdfs in five equal-width bins of $|\eta|$. This procedure also ensures that different detector regions are treated separately.

\section{Systematic uncertainties}
\label{sec:systematics}

The systematic uncertainties are obtained from data, whenever possible. The two major uncertainties for which this is not the case are cross sections and ISR, the latter being folded with the trigger efficiency curve obtained from data to produce the overall \etmiss trigger efficiency. The individual contributions are outlined below and summarised in Table~\ref{tab:rhad_ma_systsummary}.

\textbf{Theoretical cross sections}

Signal cross sections are calculated to next-to-leading order in the strong coupling constant, including the resummation of soft gluon emission at next-to-leading-logarithmic accuracy (NLO+NLL)~\cite{Beenakker:1997ut,Beenakker:2010nq,Beenakker:2011fu}. The nominal cross section and the uncertainty are taken from an envelope of cross-section predictions using different PDF sets and factorisation and renormalisation scales, as described in Ref.~\cite{Kramer:2012bx}. This prescription results in an uncertainty of $14\%$ (at 600~\GeV) rising to $24\%$ (at 1600~\GeV) and to $32\%$ (at 2000~\GeV) for gluino \rhads and marginally larger values for squark \rhads.

\textbf{Signal efficiency}

The \etmiss trigger uses only calorimeter information to calculate \etmiss, and has very low sensitivity to muons. Hence, \zmumu events can be used for calibration and to study systematic errors. To evaluate the trigger efficiency, the trigger turn-on curve is obtained by fitting the measured efficiency vs.\ \etmiss in \zmumu events, in both data and simulation. These efficiency turn-on curves are then applied to the \etmiss spectrum from simulated \rhad events. The total uncertainty is estimated from four contributions: the relative difference between the efficiencies obtained using the fitted threshold curves from \zmumu data and simulation, the differences in efficiency obtained from independent $\pm1\sigma$ variations in fit parameters relative to the unchanged turn-on-curve fit for both \zmumu data and simulation and a $10\%$ variation of the \etmiss to assess the scale uncertainty. The \etmiss trigger is estimated to contribute a total uncertainty of 2\% to the signal efficiency.

To address a possible mismodelling of ISR, and hence \etmiss in the signal events, half of the difference between the selection efficiency for the \pythia events and those reweighted with \madgraph is taken as an uncertainty in the expected signal and found to be below 14\% in all cases.

The uncertainty in the pile-up modelling in simulation is found to affect the signal efficiency by between 7\% and 1\%, decreasing as a function of the simulated \rhad mass.

The systematic uncertainty in the $\beta$ estimation is assessed by scaling the calorimeter-cell-time smearing of simulated events by $\pm$10\%, varying by $\pm1\sigma$ the parameters of the linear fit to correct the remaining $\eta$ dependence of the measured calorimeter time and by removing or doubling the cell-time correction introduced to correct the bias due to the OFA. The uncertainty is calculated as half the maximum variation in signal efficiency in all combinations divided by the average signal efficiency and is found to be between $10\%$ and $2\%$, decreasing with simulated \rhad mass.

The systematic uncertainty of the pixel $\beta\gamma$ measurement is assessed by taking into account the differences between simulation and data, the remaining variation in the reconstruction of reference masses after a run-by-run correction of an observed drift of \dedx, due to radiation sensitivity of the IBL electronics, and the stability of the \dedx-based proton mass estimate over time. The impact on the signal efficiency is obtained by applying the variations corresponding to the above-listed uncertainties independently and the overall size of these effects is found to be below $3\%$ for any simulated \rhad mass.

The uncertainty of the integrated luminosity is $\lumiUnc$, as derived following a methodology similar to that detailed in Ref.~\cite{ATLAS:lumi}, from a calibration of the luminosity scale using $x$--$y$ beam-separation scans performed in August 2015.

\textbf{Background estimation}

The uncertainty in the background estimate is evaluated by varying both the number of $|\eta|$ bins used when creating the $p$, $\beta$ and $\beta\gamma$ pdfs and the requirements on the background selection region. The nominal number of $|\eta|$ bins is varied between three and eight, while the requirements on observables are set to a \emph{medium} ($\beta < 0.975$, $\beta\gamma < 2.45$ and $60~\GeV < p < 190~\GeV$) and a \emph{tight} ($\beta < 0.95$, $\beta\gamma < 2.4$ and $70~\GeV < p < 180~\GeV$) selection. Uncertainties introduced by statistical fluctuations in the pdfs are estimated by repeating $O(100)$ times the background estimation using pdfs with Poisson variations of the content in each bin.

The correction applied to high values of calorimeter cell times measured with the OFA is found to affect the background estimate by between 5\% and 14\%. Effects arising from the \dedx measurement are assessed by using an analytical description to vary the shape of the high-ionisation tail and by changing the IBL ionisation correction by $\pm1\sigma$ and are found to be between $17\%$ and $6\%$, decreasing with simulated \rhad mass. The effect of signal contamination in the background estimation is studied by introducing the expected number of signal events into the data before building the background estimate and is found to be $10\%$ at a simulated mass of 600~\GeV, while negligible for higher masses, and is included in the overall uncertainty in the background estimate. The overall uncertainty in the background estimate is found to be $30\%$ to $43\%$, rising with simulated \rhad mass. Since the background is very small for high \rhad masses ($\geq$1400~\GeV) the relatively large uncertainty does not affect the sensitivity in this region.

\begin{table}[tbp]
  \small
  \centering
  \begin{tabular*}{\textwidth}{l| @{\extracolsep{\fill}} r}
    \hline
    \hline
    Source & Relative uncertainty~[$\pm\%$] \\
    \hline
    Theoretical uncertainty on signal                               &                14--57  \\
    \hline
    Uncertainty on signal efficiency                                &                20--16  \\
    {\footnotesize~~${}^\llcorner$~Trigger efficiency             } & {\footnotesize      2} \\
    {\footnotesize~~${}^\llcorner$~QCD uncertainty (ISR, FSR)     } & {\footnotesize     14} \\
    {\footnotesize~~${}^\llcorner$~Pile-up                        } & {\footnotesize   7--1} \\
    {\footnotesize~~${}^\llcorner$~Pixel $\beta\gamma$ measurement} & {\footnotesize   1--3} \\
    {\footnotesize~~${}^\llcorner$~Calorimeter $\beta$ measurement} & {\footnotesize  10--2} \\
    \hline
    Luminosity                                                      &                     5  \\
    \hline
    Uncertainties on background estimate                            &                30--43  \\
    \hline
    \hline
  \end{tabular*}
  \caption{Summary of all studied systematic uncertainties. Ranges indicate a dependency on the \rhad mass hypothesis (from low to high masses).}
  \label{tab:rhad_ma_systsummary}
\end{table}

\section{Results}
\label{sec:result}

\begin{table}[tbp]
  \small
  \centering
  \begin{tabular*}{\textwidth}{l|@{\extracolsep{\fill}}rr@{\extracolsep{0em}}c@{\extracolsep{0em}}l@{\extracolsep{\fill}}r@{\extracolsep{0em}}c@{\extracolsep{0em}}l@{\extracolsep{\fill}}r@{\extracolsep{0em}}c@{\extracolsep{0em}}l@{\extracolsep{\fill}}r}
    \hline
    \hline
    \rhad & Mass [\GeV] & $N_\text{sig}$&$\pm$&$\sigma_{N_\text{sig}}$ & eff.&$\pm$&$\sigma_\text{eff}$ & $N_\text{bkg}$&$\pm$&$\sigma_{N_\text{bkg}}$ & $N_\text{obs}$ \\
    \hline
    \multirow{7}{*}{Gluino}        &  600 &              3340&$\pm$&660 &           0.113&$\pm$&0.022 &               4.5&$\pm$&1.4 &     3 \\
                                   &  800 &               500&$\pm$&110 &           0.105&$\pm$&0.022 &             1.75&$\pm$&0.53 &     3 \\
                                   & 1000 &                143&$\pm$&28 &           0.137&$\pm$&0.027 &             1.23&$\pm$&0.37 &     2 \\
                                   & 1200 &              36.5&$\pm$&6.4 &           0.133&$\pm$&0.023 &             0.77&$\pm$&0.25 &     2 \\
                                   & 1400 &              12.2&$\pm$&2.2 &           0.151&$\pm$&0.028 &             0.54&$\pm$&0.19 &     2 \\
                                   & 1600 &               3.6&$\pm$&0.6 &           0.140&$\pm$&0.023 &           0.185&$\pm$&0.071 &     1 \\
                                   & 1800 &             1.00&$\pm$&0.18 &             0.11&$\pm$&0.02 &           0.138&$\pm$&0.057 &     1 \\
                                   & 2000 &           0.378&$\pm$&0.063 &             0.12&$\pm$&0.02 &           0.126&$\pm$&0.053 &     1 \\
    \hline
    \multirow{5}{*}{Bottom squark} &  600 &              36.1&$\pm$&7.7 &           0.064&$\pm$&0.014 &               4.5&$\pm$&1.4 &     3 \\
                                   &  800 &               6.6&$\pm$&1.5 &           0.073&$\pm$&0.016 &             1.75&$\pm$&0.53 &     3 \\
                                   & 1000 &             1.62&$\pm$&0.33 &           0.082&$\pm$&0.017 &             1.23&$\pm$&0.37 &     2 \\
                                   & 1200 &           0.407&$\pm$&0.077 &           0.079&$\pm$&0.015 &             0.77&$\pm$&0.25 &     2 \\
                                   & 1400 &           0.122&$\pm$&0.024 &           0.082&$\pm$&0.016 &             0.54&$\pm$&0.19 &     2 \\
    \hline
    \multirow{5}{*}{Top squark}    &  600 &              47.5&$\pm$&9.5 &           0.085&$\pm$&0.017 &               4.5&$\pm$&1.4 &     3 \\
                                   &  800 &              10.7&$\pm$&2.3 &           0.118&$\pm$&0.025 &             1.75&$\pm$&0.53 &     3 \\
                                   & 1000 &             2.70&$\pm$&0.52 &           0.137&$\pm$&0.026 &             1.23&$\pm$&0.37 &     2 \\
                                   & 1200 &             0.72&$\pm$&0.13 &           0.141&$\pm$&0.025 &             0.77&$\pm$&0.25 &     2 \\
                                   & 1400 &           0.216&$\pm$&0.039 &           0.146&$\pm$&0.027 &             0.54&$\pm$&0.19 &     2 \\
    \hline
    \hline
  \end{tabular*}
  \caption{Expected signal yield ($N_\text{sig}$) and efficiency (eff.), estimated background ($N_\text{bkg}$) and observed number of events in data ($N_\text{obs}$) for the full mass range after the final selection using \lumi of data. The stated uncertainties include both the statistical and systematic contribution.}
  \label{tab:rhad_ma_finalnumbers}
\end{table}

\begin{figure}[tbp]
  \centering
  \includegraphics[width=.58\textwidth]{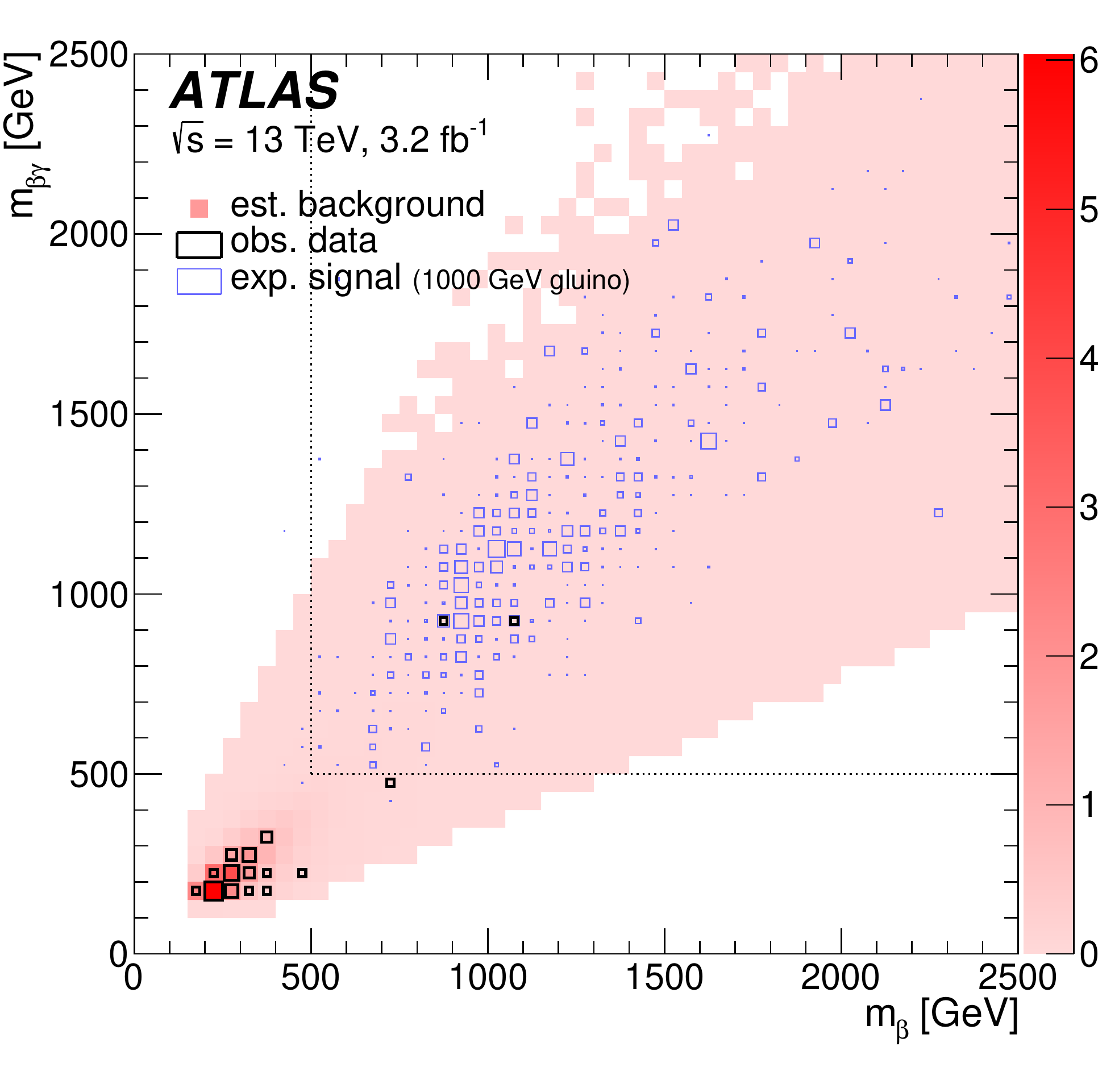}
  \caption{Data (bold boxes) and background estimates (colour fill) for $m_{\beta}$ vs.\ $m_{\beta\gamma}$ for the gluino \rhad search (1000~\GeV). The blue thin-line boxes illustrate the expected signal (on top of background) for the given \rhad mass hypothesis. The black dashed vertical/horizontal lines at 500~\GeV show the mass selection (signal region in the top-right). Two events pass this selection. (For interpretation of the references to colour in this figure legend, the reader is referred to the web version of this article.)}
  \label{fig:rhadmsagnosr1000}
\end{figure}

\begin{figure}[tbp]
  \centering
  \includegraphics[width=.48\textwidth]{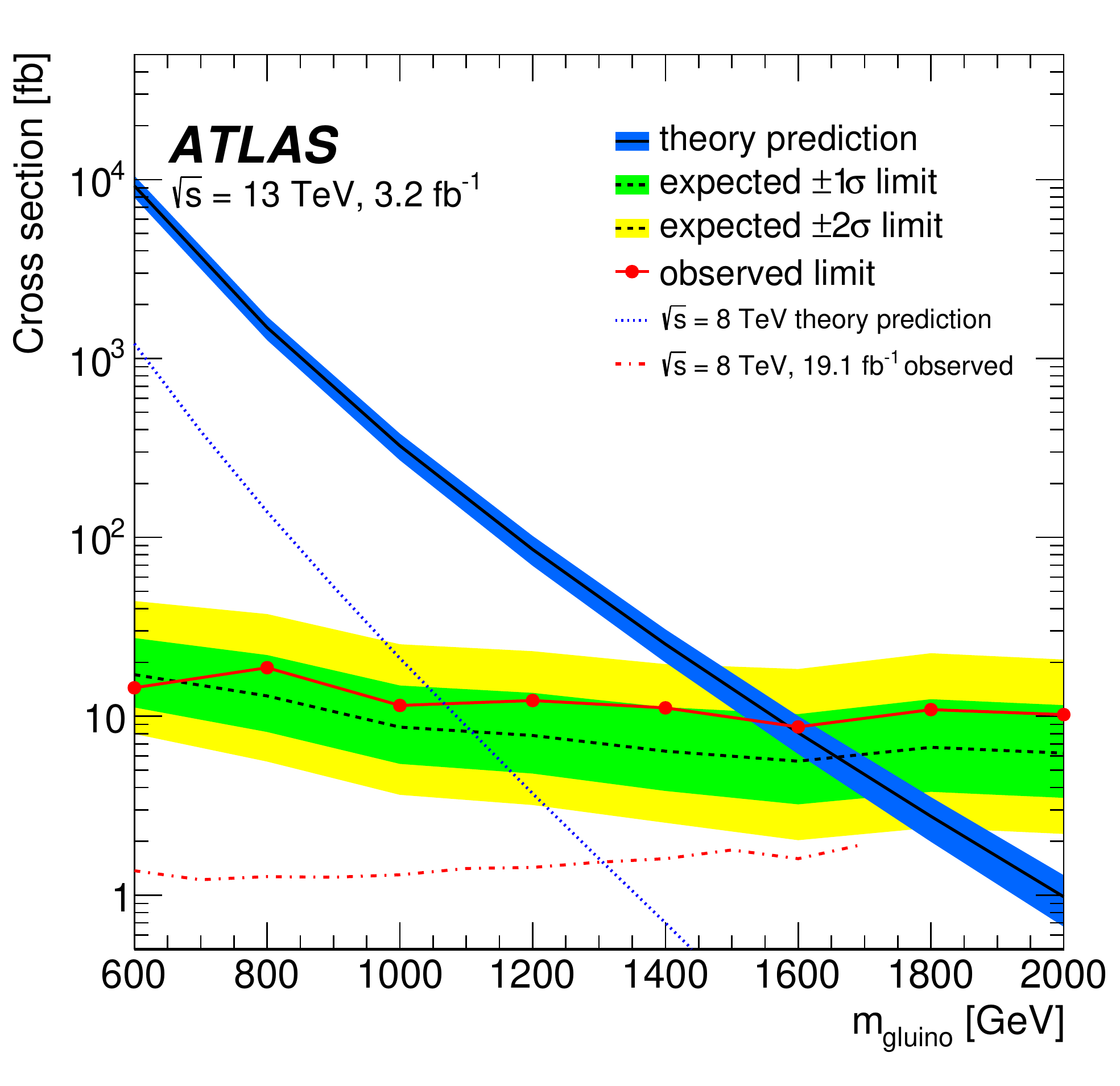}\\
  \includegraphics[width=.48\textwidth]{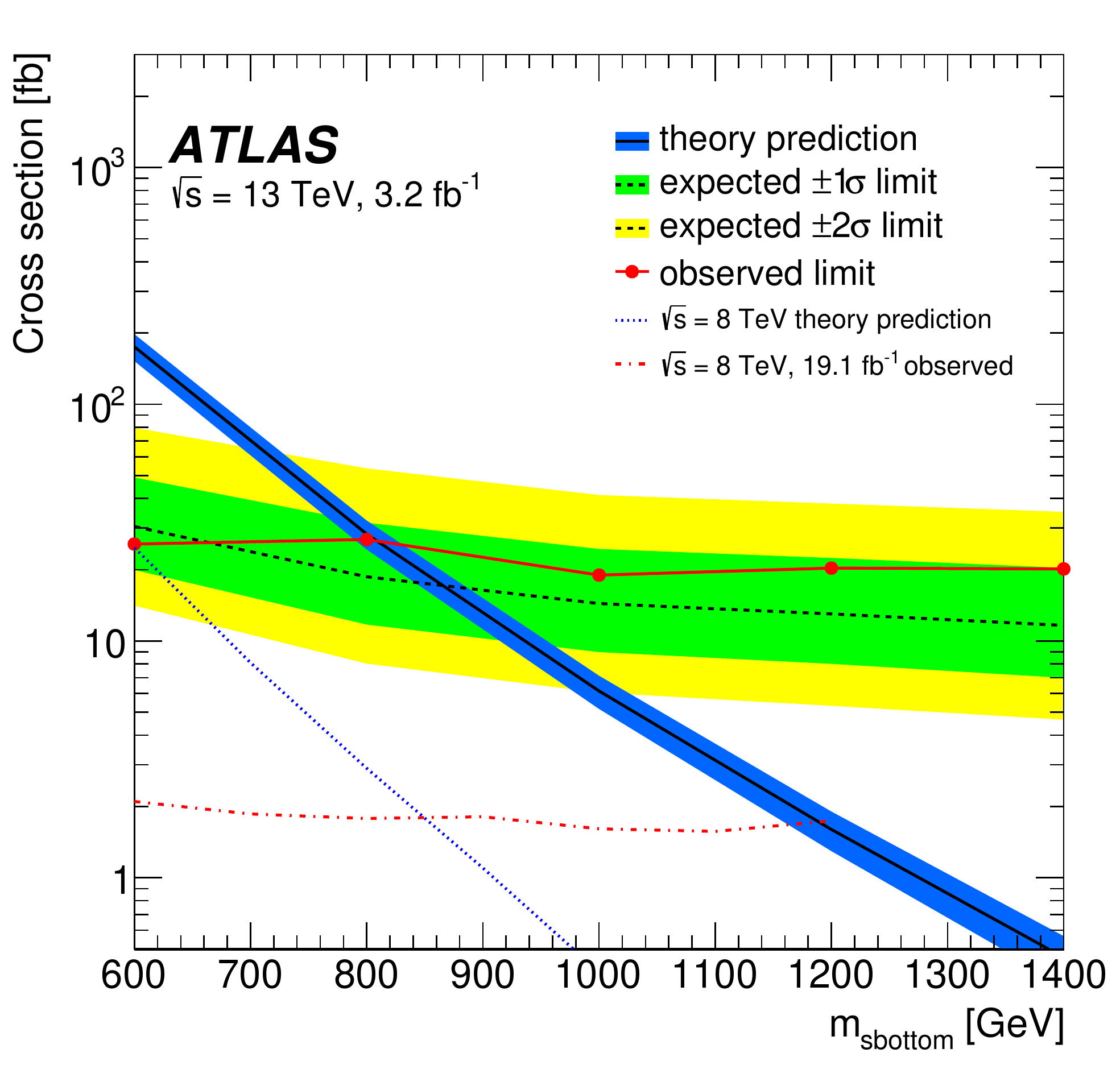}
  \includegraphics[width=.48\textwidth]{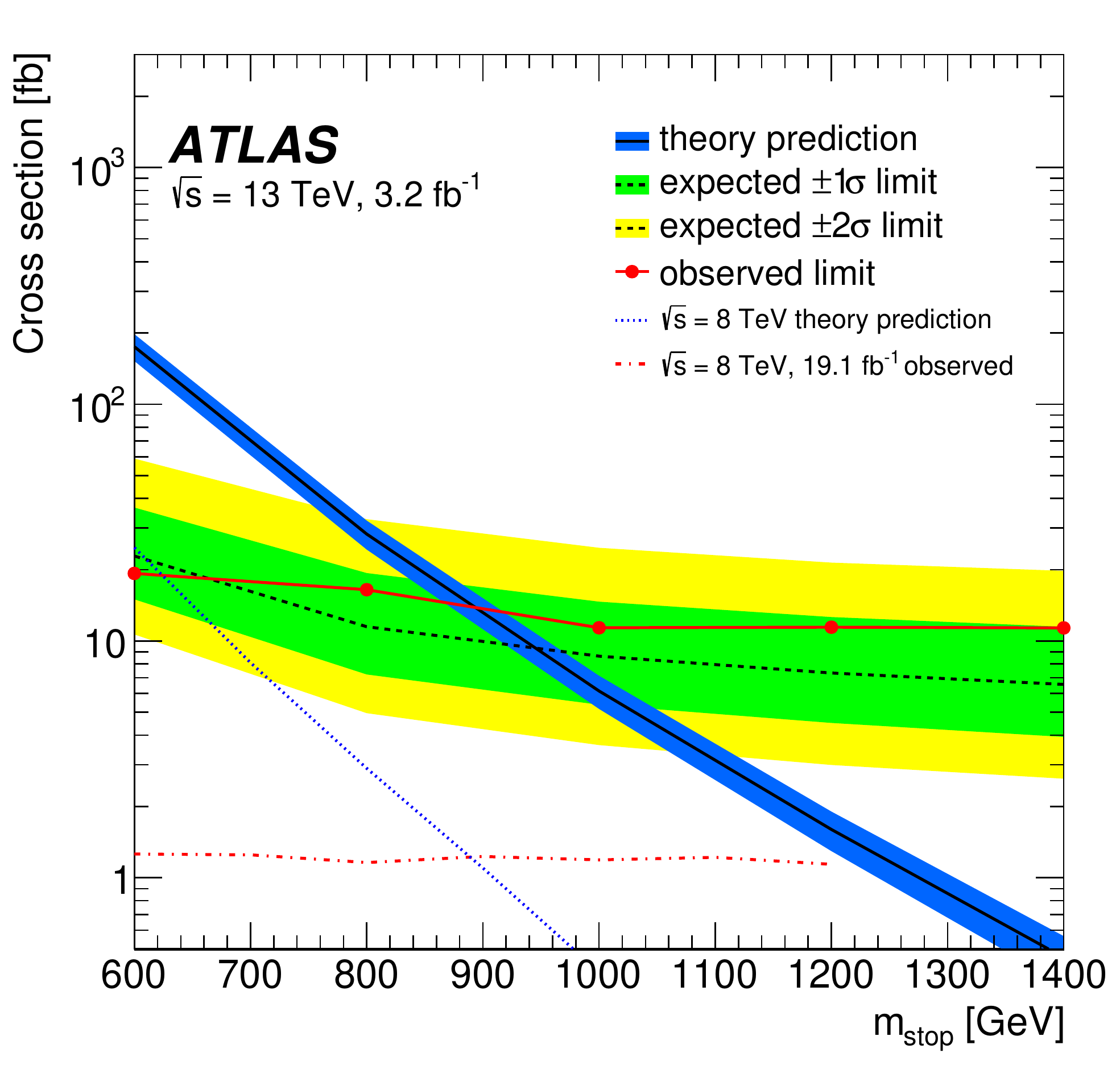}
  \caption{Expected (dashed black line) and observed (solid red line) 95\% CL upper limits on the cross section as a function of mass for the production of long-lived gluino (top), bottom-squark (bottom-left) and top-squark (bottom-right) \rhads. The theory prediction along with its $\pm1\sigma$ uncertainty is show as a black line and a blue band, respectively. The observed 8~\TeV Run-1 limit and theory prediction~\cite{ATLAS:2014fka} are shown in dash-dotted and dotted lines, respectively. (For interpretation of the references to colour in this figure legend, the reader is referred to the web version of this article.)}
  \label{fig:limitrhadMA}
\end{figure}

The resulting mass distributions of events for the 1000~\GeV gluino \rhad mass hypothesis can be seen in Figure~\ref{fig:rhadmsagnosr1000}. Two events with masses above 500~\GeV pass the event selection for the 1000~\GeV mass hypothesis, while only one of these events passes the event selection for the 1600~\GeV mass hypothesis. However, as can be seen in Table~\ref{tab:rhad_ma_finalnumbers}, at no point in the examined mass range does this search exhibit any statistically significant excess of events above the expected background, which is $1.23\pm0.37$ and $0.185\pm0.071$ for the two above-mentioned mass hypotheses, respectively. Therefore, 95\% CL upper limits are placed on the \rhad production cross section, as shown in Figure~\ref{fig:limitrhadMA}. These limits are obtained from the expected signal and the estimated background in the signal region and using a one-bin counting experiment applying the $CL_s$ prescription~\cite{Read:2000ru}.

Given the predicted theoretical cross sections, also shown in Figure~\ref{fig:limitrhadMA}, the cross-section limits are translated into lower limits on \rhad masses. Expected lower limits at 95\% CL on the \rhad masses of \limitExpGluinoStable, \limitExpSbottomStable and \limitExpStopStable for the production of long-lived gluino, bottom-squark and top-squark \rhads are derived, respectively. Corresponding observed lower mass limits at 95\% CL for gluino, bottom-squark and top-squark \rhads are found to be \limitGluinoStable, \limitSbottomStable and \limitStopStable, respectively.

For comparison, the corresponding ATLAS Run-1 8~\TeV lower limits at 95\% CL on the mass of gluino, bottom-squark and top-squark \rhads~\cite{ATLAS:2014fka} are also shown in Figure~\ref{fig:limitrhadMA}.

\FloatBarrier

\section{Conclusion}
\label{sec:conclusion}

A search for heavy long-lived particles in the form of composite colourless states of squarks or gluinos together with SM quarks and gluons, called \rhads, and taking advantage of both ionisation and time-of-flight measurements is presented in this Letter. The search uses 3.2 fb$^{-1}$ of $pp$ collisions at $\sqrt{s}$ = 13 \TeV collected by the ATLAS experiment at the LHC. No statistically significant excess of events above the expected background is found for any \rhad mass hypothesis. Long-lived \rhads containing a gluino, bottom or top squark are excluded at 95\% CL for masses up to \limitGluinoStable, \limitSbottomStable and \limitStopStable, respectively. These results substantially extend previous ATLAS and CMS limits from 8~\TeV Run-1 data in case of gluino \rhads and are complementary to searches for SUSY particles which decay promptly.

\section*{Acknowledgements}

\input{Acknowledgements}

%



\printbibliography


\newpage
\FloatBarrier
\input{atlas_authlist}


%
\end{document}

%% file: Acknowledgements.tex

We thank CERN for the very successful operation of the LHC, as well as the
support staff from our institutions without whom ATLAS could not be
operated efficiently.

We acknowledge the support of ANPCyT, Argentina; YerPhI, Armenia; ARC, Australia; BMWFW and FWF, Austria; ANAS, Azerbaijan; SSTC, Belarus; CNPq and FAPESP, Brazil; NSERC, NRC and CFI, Canada; CERN; CONICYT, Chile; CAS, MOST and NSFC, China; COLCIENCIAS, Colombia; MSMT CR, MPO CR and VSC CR, Czech Republic; DNRF and DNSRC, Denmark; IN2P3-CNRS, CEA-DSM/IRFU, France; GNSF, Georgia; BMBF, HGF, and MPG, Germany; GSRT, Greece; RGC, Hong Kong SAR, China; ISF, I-CORE and Benoziyo Center, Israel; INFN, Italy; MEXT and JSPS, Japan; CNRST, Morocco; FOM and NWO, Netherlands; RCN, Norway; MNiSW and NCN, Poland; FCT, Portugal; MNE/IFA, Romania; MES of Russia and NRC KI, Russian Federation; JINR; MESTD, Serbia; MSSR, Slovakia; ARRS and MIZ\v{S}, Slovenia; DST/NRF, South Africa; MINECO, Spain; SRC and Wallenberg Foundation, Sweden; SERI, SNSF and Cantons of Bern and Geneva, Switzerland; MOST, Taiwan; TAEK, Turkey; STFC, United Kingdom; DOE and NSF, United States of America. In addition, individual groups and members have received support from BCKDF, the Canada Council, CANARIE, CRC, Compute Canada, FQRNT, and the Ontario Innovation Trust, Canada; EPLANET, ERC, FP7, Horizon 2020 and Marie Sk{\l}odowska-Curie Actions, European Union; Investissements d'Avenir Labex and Idex, ANR, R{\'e}gion Auvergne and Fondation Partager le Savoir, France; DFG and AvH Foundation, Germany; Herakleitos, Thales and Aristeia programmes co-financed by EU-ESF and the Greek NSRF; BSF, GIF and Minerva, Israel; BRF, Norway; Generalitat de Catalunya, Generalitat Valenciana, Spain; the Royal Society and Leverhulme Trust, United Kingdom.

The crucial computing support from all WLCG partners is acknowledged
gratefully, in particular from CERN and the ATLAS Tier-1 facilities at
TRIUMF (Canada), NDGF (Denmark, Norway, Sweden), CC-IN2P3 (France),
KIT/GridKA (Germany), INFN-CNAF (Italy), NL-T1 (Netherlands), PIC (Spain),
ASGC (Taiwan), RAL (UK) and BNL (USA) and in the Tier-2 facilities
worldwide.

%% file: atlas_authlist.tex
\begin{flushleft}
{\Large The ATLAS Collaboration}

\bigskip

M.~Aaboud$^\textrm{\scriptsize 135d}$,
G.~Aad$^\textrm{\scriptsize 86}$,
B.~Abbott$^\textrm{\scriptsize 113}$,
J.~Abdallah$^\textrm{\scriptsize 64}$,
O.~Abdinov$^\textrm{\scriptsize 12}$,
B.~Abeloos$^\textrm{\scriptsize 117}$,
R.~Aben$^\textrm{\scriptsize 107}$,
O.S.~AbouZeid$^\textrm{\scriptsize 137}$,
N.L.~Abraham$^\textrm{\scriptsize 149}$,
H.~Abramowicz$^\textrm{\scriptsize 153}$
% ...
\bigskip
\\
% ...
$^{12}$ Institute of Physics, Azerbaijan Academy of Sciences, Baku, Azerbaijan\\
% ...
$^{64}$ University of Iowa, Iowa City IA, United States of America\\
% ...
$^{86}$ CPPM, Aix-Marseille Universit{\'e} and CNRS/IN2P3, Marseille, France\\
% ...
$^{107}$ Nikhef National Institute for Subatomic Physics and University of Amsterdam, Amsterdam, Netherlands\\
% ...
$^{113}$ Homer L. Dodge Department of Physics and Astronomy, University of Oklahoma, Norman OK, United States of America\\
% ...
$^{117}$ LAL, Univ. Paris-Sud, CNRS/IN2P3, Universit{\'e} Paris-Saclay, Orsay, France\\
% ...
$^{135}$ $^{(a)}$ Facult{\'e} des Sciences Ain Chock, R{\'e}seau Universitaire de Physique des Hautes Energies - Universit{\'e} Hassan II, Casablanca; $^{(b)}$ Centre National de l'Energie des Sciences Techniques Nucleaires, Rabat; $^{(c)}$ Facult{\'e} des Sciences Semlalia, Universit{\'e} Cadi Ayyad, LPHEA-Marrakech; $^{(d)}$ Facult{\'e} des Sciences, Universit{\'e} Mohamed Premier and LPTPM, Oujda; $^{(e)}$ Facult{\'e} des sciences, Universit{\'e} Mohammed V, Rabat, Morocco\\
% ...
$^{137}$ Santa Cruz Institute for Particle Physics, University of California Santa Cruz, Santa Cruz CA, United States of America\\
% ...
$^{149}$ Department of Physics and Astronomy, University of Sussex, Brighton, United Kingdom\\
% ...
$^{153}$ Raymond and Beverly Sackler School of Physics and Astronomy, Tel Aviv University, Tel Aviv, Israel\\
% ...
$^{a}$ Also at Department of Physics, King's College London, London, United Kingdom\\
$^{b}$ Also at Institute of Physics, Azerbaijan Academy of Sciences, Baku, Azerbaijan\\
$^{c}$ Also at Novosibirsk State University, Novosibirsk, Russia\\
$^{d}$ Also at TRIUMF, Vancouver BC, Canada\\
$^{e}$ Also at Department of Physics {\&} Astronomy, University of Louisville, Louisville, KY, United States of America\\
% ...
\end{flushleft}
